\documentclass[12pt]{article}\pdfoutput=1

\usepackage[intlimits]{amsmath}
\usepackage{amsfonts}
\usepackage{dsfont}
\usepackage{bbm}
\usepackage[retainorgcmds]{IEEEtrantools}
\usepackage[colorlinks=true, linktoc=page, citecolor=blue, urlcolor=blue]{hyperref}
\usepackage{verbatim}
\usepackage[bulletsep]{collref}

\newcommand{\CP}{{\mathbb{CP}}}
\newcommand{\x}{{\textrm{x}}}
\newcommand{\z}{{\textrm{z}}}
\newcommand{\M}{\mathcal{M}}
\newcommand{\T}{\mathcal{T}}
\newcommand{\U}{\mathbb{U}}
\newcommand{\Q}{\mathfrak{Q}}
\newcommand{\eq}{\stackrel{!}{=}}
\newcommand{\st}{\text{st}}
\newcommand{\Pexp}{\overrightarrow{\rm P}\exp}
\newcommand{\expP}{\overleftarrow{\rm P}\exp}

\makeatletter \@addtoreset{equation}{section} \makeatother

\hypersetup{pdfstartview={XYZ null null 0.75}}
\begin{document}

\renewcommand{\thefootnote}{\fnsymbol{footnote}}
\setcounter{footnote}{0}
\begin{center}
{\Large\textbf{\mathversion{bold} \textbf{String integrability of the ABJM defect}\\[12pt]}\par}
\vspace{0.6cm}

\textrm{Georgios~Linardopoulos}
\vspace{8mm}

\textit{Wigner Research Centre for Physics \\ Konkoly-Thege Mikl{ó}s \'{u}t 29-33, 1121 Budapest, Hungary} \\[6pt]
\vspace{0.2cm}
\texttt{george.linardopoulos@wigner.hu}
\vspace{1cm}

\textbf{Abstract}\vspace{3mm}

\begin{minipage}{13cm}
\noindent ABJM theory in the presence of a half-BPS domain wall is dual to the D2-D4 probe brane system with nonzero worldvolume flux. The ABJM domain wall was recently shown to be integrable to lowest order in perturbation theory and bond dimension. In the present paper we show that the string theory dual of this system is integrable, namely that the string boundary conditions on the probe D4-brane preserve the integrability of the Green-Schwarz sigma model. Our result suggests that the ABJM domain wall is integrable to all loop orders and for any value of the bond dimension.
\end{minipage}
\end{center} \vspace{3mm}

\setcounter{page}{1}
\renewcommand{\thefootnote}{\arabic{footnote}}
\setcounter{footnote}{0}
\flushbottom

\newpage\section[Introduction]{Introduction}
\noindent The IIA/ABJM correspondence is an AdS$_4$/CFT$_3$ type of holographic duality that arises in the weak-coupling limit of the M/ABJM correspondence \cite{ABJM08}. The duality postulates the equivalence of 3-dimensional ABJM theory and type IIA superstring theory in AdS$_4 \times \CP^3$:
\begin{IEEEeqnarray}{c}
\begin{array}{c} \mathcal{N} = 6, \ U\left(N\right)_k \times U\left(N\right)_{-k} \\[5pt] \text{super Chern-Simons theory} \\[5pt] k^5 \gg N \rightarrow \infty \ \& \ \lambda \equiv N/k \end{array} \ \Leftrightarrow \ \begin{array}{c} \text{Type IIA string theory on} \\[5pt] \text{AdS}_4 \times \CP^3 \text{ with } N \text{ units of flux} \\[5pt] \text{in AdS}_4 \text{ and } k \text{ units in }\CP^3. \end{array} \nonumber
\end{IEEEeqnarray}
\indent The AdS$_4$/CFT$_3$ duality has many common features with the AdS$_5$/CFT$_4$ duality.\footnote{Hereafter AdS$_5$/CFT$_4$ denotes the duality between type IIB string theory on AdS$_5\times\text{S}^5$ and $\mathcal{N}=4$ super Yang Mills theory (SYM). Likewise, AdS$_4$/CFT$_3$ denotes the duality between type IIA string theory on AdS$_4\times\CP^3$ and ABJM theory. More about the integrability of these theories can be found in the reviews \cite{KristjansenStaudacherTseytlin09, Beisertetal12}. \cite{Klose10} covers AdS$_4$/CFT$_3$ integrability.} Besides holography, these dualities also share the property of planar integrability. In the former case planar integrability was established at both strong \cite{ArutyunovFrolov08, Stefanski08} and weak coupling \cite{MinahanZarembo09}, following earlier progress in the latter \cite{BenaPolchinskiRoiban03, MinahanZarembo03}. Integrability has important consequences for the solution of the underlying theories as it allows for a non-perturbative treatment of their observables by the spectral curve method \cite{GromovKazakovLeurentVolin13, CavagliaFioravantiGromovTateo14}. Another common property of the two dualities is that they both afford integrable deformations by probe branes. \\
\indent Integrable deformations of holographic theories are interesting because they lead to more and more realistic physical models that break many of the original symmetries and supersymmetries, but preserve integrability with all of its powerful features.\footnote{An early account of integrable deformations can be found in \cite{Zoubos10}.} Deforming holographic dualities by introducing probe branes on their string theory side was initiated long ago by Karch and Randall \cite{KarchRandall01a,KarchRandall01b}. Even if the undeformed theory is integrable, the probe brane that hosts the string boundary conditions, can break its integrability. On the other hand, integrable branes always lead to the conservation of an infinite number of classical charges. \\
\indent The dual gauge theories of holographic dualities that have been deformed by probe branes are often defect conformal field theories (dCFTs). These are again deformations of the duality where the action of the (bulk) gauge theory is coupled to the action of a lower-dimensional (boundary) gauge theory \cite{DeWolfeFreedmanOoguri01, DeWolfeMann04}. Defect CFTs can be studied by means of domain walls, i.e.\ classical solutions of the bulk equations of motion \cite{ConstableMyersTafjord99, ConstableMyersTafjord01a} with or without Nahm poles \cite{NagasakiTanidaYamaguchi11, NagasakiYamaguchi12, KristjansenSemenoffYoung12b}. When the undeformed gauge theory is integrable, the corresponding domain wall can be studied with integrability methods. Specifically one can compute correlators as overlaps of boundary states (matrix product states of various bond dimensions) and the Bethe eigenstates of the integrable spin chain that describes dilatations in the undeformed theory (e.g.\ 1-point functions at tree level \cite{deLeeuwKristjansenZarembo15, Buhl-MortensenLeeuwKristjansenZarembo15, deLeeuwKristjansenMori16, deLeeuwKristjansenLinardopoulos16, deLeeuwKristjansenLinardopoulos18a, deLeeuwGomborKristjansenLinardopoulosPozsgay19, KristjansenMullerZarembo20a, KristjansenMullerZarembo20b} and loop-order \cite{Buhl-MortensenLeeuwIpsenKristjansenWilhelm16a, Buhl-MortensenLeeuwIpsenKristjansenWilhelm16c, Buhl-MortensenLeeuwIpsenKristjansenWilhelm17a, GimenezGrauKristjansenVolkWilhelm19, GomborBajnok20a, GomborBajnok20b, KristjansenMullerZarembo21}, 2-point functions \cite{deLeeuwIpsenKristjansenVardinghusWilhelm17, Widen17} etc., see the reviews \cite{deLeeuwIpsenKristjansenWilhelm17, deLeeuw19, Linardopoulos20}). \\
\indent Once more, the defect boundary conditions (such as \cite{GaiottoWitten08a}) can break the integrability of the undeformed theory. A powerful criterion for the integrability of domain walls was developed in \cite{PiroliPozsgayVernier17} by exploiting an intriguing link between defect quantum field theories and quantum quenches \cite{GhoshalZamolodchikov93}. Integrable domain walls, (1) have the parity-odd charges of the spin chain annihilate the boundary state, (2) have nontrivial overlaps only when the Bethe roots come in pairs of opposite signs, and (3) are solvable in the sense that overlaps are given by closed-form expressions. Ideally, this list would also include integrability on the string theory side of holographic domain walls. \\
\indent Several domain wall versions of $\mathcal{N} = 4$ SYM have been shown to satisfy all of the above integrable quench criteria. One of them is the holographic dual of a probe D5-brane that lives on the string theory side of the AdS$_5$/CFT$_4$ duality. The D3-D5 intersection consists of an AdS$_4\times\text{S}^2$ shaped D5-brane inside AdS$_5\times\text{S}^5$, with $k$ units of magnetic flux through S$^2$. The flux number $k$ is also equal to the number of D3-branes that terminate upon the probe D5-brane, as well as to the bond dimension of the domain wall boundary state. The integrability of the D5-brane for any value of $k$ was proven in \cite{LinardopoulosZarembo21}. Dekel and Oz had previously shown integrability in the $k = 0$ case \cite{DekelOz11b}. \\
\indent In this work we focus on the domain wall version of ABJM theory, the holographic dual of which is obtained by adding a probe D4-brane on the string theory side of the AdS$_4$/CFT$_3$ duality. The resulting D2-D4 probe brane system is made up from a AdS$_3\times\CP^1$ shaped D4-brane inside AdS$_4\times\CP^3$, with $Q$ units of magnetic flux through $\CP^1$. Again $Q$ is related to the number of D2-branes that terminate on one side of the probe D4-brane and the bond dimension of the domain wall boundary state. \\
\indent Very recently, strong evidence suggesting the planar integrability of the ABJM domain wall was found \cite{KristjansenVuZarembo21}. In particular, the authors of \cite{KristjansenVuZarembo21} were able to show that the boundary state of the ABJM domain wall satisfies the integrable quench criteria (2) and (3), to lowest order in perturbation theory and bond dimension. The scalar sector of their closed-form determinant formula agrees with an overlap formula for the alternating $\mathfrak{su}(4)$ spin chain that was derived in \cite{Gombor21} and proposed to be relevant for ABJM theory. \\
\indent The main aim of the present paper will be to show that the string theory dual of the ABJM domain wall is integrable for all values of the flux $Q$, complementing the earlier result of \cite{DekelOz11b} which showed integrability in the zero-flux case $Q = 0$. To generate an infinite family of classically conserved charges for the D4-brane (and any integrable field theory with a boundary) we need to construct its double-row monodromy matrix $\T$ \cite{Sklyanin87a}. $\T$ is formed by sandwiching a reflection matrix $\U$ between two monodromy matrices $\M$. The conserved charges are then obtained by expanding the double-row monodromy matrix around appropriate values of the spectral parameter $\x$. The latter is always possible when the double-row monodromy matrix is conserved on the brane so that the D-brane integrability condition can be formulated in terms of an equation that should be obeyed by the reflection matrix $\U$. \\
\indent As a first step, we identify the integrable reflection matrix of the probe D4-brane. The matrix depends on both the spectral parameter $\x$ and the string embedding coordinates, in contradistinction with the zero-flux reflection matrices of Dekel and Oz \cite{DekelOz11b} which had constant matrix elements. Thereafter we expand the monodromy matrix to obtain an infinite family of conserved quantities. We further show that the unbroken symmetry group is $SO(2,2)\times SU(2)\times SU(2)\times U(1)$, which is consistent with the AdS$_4\times\mathbb{CP}^1$ geometry of the D4-brane. Our analysis also confirms that the probe D4-brane is half-BPS, preserving 12 out of the initial 24 supercharges \cite{ChandrasekharPanda09}. \\
\indent The paper is organized as follows. We start (in section \ref{Section:StringSigmaModel}) by revisiting the string sigma model in AdS$_4\times\CP^3$ and the integrability conditions for probe D-branes in it. In \S\ref{Section:D2D4intersection} we briefly review the D2-D4 probe brane system and formulate the string boundary conditions on the D4-brane. In section \ref{Section:D4braneIntegrability} we specify the integrable reflection matrices of the D4-brane in both AdS$_4$ and $\CP^3$. In section \ref{Section:ConservedCharges} we expand the double-row monodromy matrix and obtain an infinite tower of classically conserved charges. The charges and supercharges are subsequently shown to fully agree with the symmetries of the probe D4-brane. In \S\ref{Section:Conclusions} we discuss some interesting extensions of our work.
\section[Integrable branes in AdS$_4\times\CP^3$]{Integrable branes in AdS$_4\times\CP^3$ \label{Section:StringSigmaModel}}
Superstring theory in AdS$_4\times\CP^3$ is described by a Green-Schwarz sigma model on the supercoset space $OSP(2,2|6)/SO(3,1)\times U(3)$ \cite{ArutyunovFrolov08, Stefanski08}:
\begin{IEEEeqnarray}{c}
S = -\frac{\ell^2}{4\pi\alpha'} \int \text{str}\left[J^{(2)}\wedge \star J^{(2)}\right] + \text{str}\left[J^{(1)} \wedge J^{(3)}\right], \quad \frac{\ell^2}{2\pi\alpha'} = \sqrt{\frac{\lambda}{2}}, \qquad \label{StringActionOSP226}
\end{IEEEeqnarray}
where $\lambda = N/k$ is the 't Hooft coupling ($k \in \mathds{Z}$ is the Chern-Simons level of ABJM) and $\ell$ is the radius of AdS$_4$. The $\mathds{Z}_4$ decomposition of the moving-frame current $J$ reads
\begin{IEEEeqnarray}{c}
J \equiv \mathfrak{g}^{-1}d\mathfrak{g} = J^{(0)} + J^{(1)} + J^{(2)} + J^{(3)}, \qquad \Omega\left[J^{(n)}\right] = i^n J^{(n)},
\end{IEEEeqnarray}
where $\mathfrak{g}$ is an $OSP(2,2|6)$ element. The $\mathds{Z}_4$ automorphism of $\mathfrak{osp}\left(2,2|6\right)$, $\Omega$ is given by
\begin{IEEEeqnarray}{c}
\Omega\left(M\right) = -\mathcal{K}M^{\st}\mathcal{K}^{-1}, \qquad
\end{IEEEeqnarray}
where $M$ is also an $OSP(2,2|6)$ element and the matrix $\mathcal{K}$ is defined as
\begin{IEEEeqnarray}{c}
\mathcal{K} = \left(\begin{array}{cc} K_4 & 0 \\ 0 & -K_6\end{array}\right), \qquad K_4 \equiv \gamma_{12}, \qquad K_6 \equiv I_3 \otimes \left(i\sigma_{2}\right). \qquad
\end{IEEEeqnarray}
See appendix \ref{Appendix:GammaMatrices} for the definition of the Dirac matrices. Their exact form is not generally required for what follows, although it might be useful in checking the various results explicitly. The moving-frame current $J$ has a vanishing curvature:
\begin{IEEEeqnarray}{c}
dJ + J \wedge J = 0,
\end{IEEEeqnarray}
while the equations of motion that follow from the action \eqref{StringActionOSP226} afford the Lax representation
\begin{IEEEeqnarray}{c}
dL + L \wedge L = 0.
\end{IEEEeqnarray}
The flat Lax connection $L$ is given by\footnote{For simplicity, we will frequently omit the dependencies of the currents and Lax connections on the worldsheet coordinates $\sigma,\tau$.}
\begin{IEEEeqnarray}{c}
L\left(\x\right) = J^{(0)} + \frac{{\x}^2+1}{{\x}^2-1}\,J^{(2)} - \frac{2{\x}}{{\x}^2-1} \star J^{(2)} + \z \, J^{(1)} + \frac{1}{\z} \, J^{(3)}, \qquad \label{LaxConnectionMovingFrame}
\end{IEEEeqnarray}
where
\begin{IEEEeqnarray}{c}
\z \equiv \sqrt{\frac{\x+1}{\x-1}}.
\end{IEEEeqnarray}
Alternatively, the Lax connection can be expressed in terms of the fixed-frame current:
\begin{IEEEeqnarray}{c}
j \equiv \mathfrak{g} J \mathfrak{g}^{-1} = d\mathfrak{g}\,\mathfrak{g}^{-1} = j^{(0)} + j^{(1)} + j^{(2)} + j^{(3)}, \qquad j^{(n)} \equiv \mathfrak{g} J^{(n)} \mathfrak{g}^{-1}, \qquad
\end{IEEEeqnarray}
which is also flat
\begin{IEEEeqnarray}{c}
dj - j \wedge j = 0,
\end{IEEEeqnarray}
making the corresponding Lax connection assume the form:
\begin{IEEEeqnarray}{c}
a\left(\x\right) = \frac{2}{\x^2 - 1} \left(j^{(2)} - \x \star j^{(2)}\right) + \left({\z} - 1\right) \, j^{(1)} + \left(\frac{1}{{\z}} - 1\right) \, j^{(3)}. \qquad \label{LaxConnectionFixedFrame}
\end{IEEEeqnarray}
The fixed-frame Lax connection \eqref{LaxConnectionFixedFrame} also satisfies a flatness condition:
\begin{IEEEeqnarray}{c}
da + a \wedge a = 0.
\end{IEEEeqnarray}
\paragraph{Closed string integrability} The existence of a flat Lax connection (either in the moving or in the fixed frame) for the 2-dimensional field theory \eqref{StringActionOSP226}, automatically implies its classical integrability. The standard way to see how the flatness of the Lax connection gives rise to an infinite set of motion integrals (which in turn provide exact solutions to the string equations of motion) is via the monodromy matrix $\M$. Due to the flatness of the Lax connection, the monodromy matrix, which is given by the Wilson line (or holonomy) of the Lax connection (see \eqref{SingleRowMonodromyMatrix} below), has a $\tau$-independent spectrum. The classically conserved charges then follow by expanding $\M$ around appropriate values of the spectral parameter $\x$.\footnote{Details can be found in many relevant textbooks, e.g.\ \cite{BabelonBernardTalon03}} Because type II string theories only describe closed strings, periodic boundary conditions are integrable for the string sigma model \eqref{StringActionOSP226}.
\paragraph{Open string integrability} Now let us take up the issue of open string integrability. We start off by being completely general; we will specify the integrability conditions for open strings in AdS$_4\times\CP^3$ by the end of this section. Open string motion can be described by attaching the string worldsheet on a D-brane. The D-brane then acts as a boundary that hosts one or both endpoints of the string. Commonly, the brane cuts the string worldsheet at a constant-$\sigma$ section. We thus want to examine the classical integrability of a string sigma model like \eqref{StringActionOSP226} in the presence of a probe D-brane at $\sigma = 0$. In other words, we ask whether the probe D-brane that generates the string boundary conditions is integrable i.e.\ it gives rise to integrable string motion. Note that closed strings correspond to periodic boundary conditions and can also be accommodated in this formalism. Following \cite{LinardopoulosZarembo21}, we define the double-row monodromy matrix,
\begin{IEEEeqnarray}{c}
\T\left(\tau;\x\right) = \M^{\st}\left(\tau;-\x\right)\cdot \U \left(\tau;\x\right)\cdot \M\left(\tau;\x\right), \label{DoubleRowMonodromyMatrix1}
\end{IEEEeqnarray}
where $\U\left(\tau;\x\right)$ is the reflection matrix at $\sigma = 0$, and the (single-row) monodromy matrix $\M\left(\tau;\x\right)$ is given by
\begin{equation}
\M\left(\tau;\x\right) = \mathfrak{g}\left(\tau,0\right) \cdot \Pexp\left(\int_{0}^{\infty} ds \, L_{\sigma}\left(s,\tau ; \x\right)\right). \label{SingleRowMonodromyMatrix}
\end{equation}
The most general reflection matrix $\U\left(\tau;\x\right)$ is dynamical, i.e.\ it can depend not only on the spectral parameter $\x$ but also implicitly on time by means of an explicit dependence on the string embedding coordinates at the $\sigma = 0$ boundary. \\
\indent An infinite family of conserved charges can be generated by demanding the double-row monodromy matrix $\T\left(\tau;\x\right)$ to be independent of the worldsheet time $\tau$, i.e.\
\begin{IEEEeqnarray}{c}
\dot{\T}\left(\tau;\x\right) \eq 0.\footnote{The exclamation mark over the equals sign $\eq$ will henceforth denote an equality that holds on the boundary, e.g.\ a boundary condition.}
\end{IEEEeqnarray}
It follows that a given set of string boundary conditions is classically integrable if it satisfies the equation
\begin{IEEEeqnarray}{c}
\dot{\U}\left(\x\right) \eq a_{\tau}^{\st}\left(-\x\right)\U\left(\x\right) + \U\left(\x\right)a_{\tau}\left(\x\right), \label{IntegrableBCs1}
\end{IEEEeqnarray}
on the $\sigma = 0$ boundary. Plugging \eqref{LaxConnectionFixedFrame} into \eqref{IntegrableBCs1} we are led to the integrability condition
\begin{IEEEeqnarray}{ll}
\dot{\U} \eq & \frac{2}{\x^2 - 1} \cdot \left\{j^{(2)\,\st}_{\tau} \, \U + \U \, j^{(2)}_{\tau}\right\} + \frac{2\x}{\x^2 - 1} \cdot \left\{j^{(2)\,\st}_{\sigma} \, \U - \U \, j^{(2)}_{\sigma}\right\} + \nonumber \\[6pt]
& + \left({\z} - 1\right) \cdot \left\{j^{(3)\,\st}_{\tau} \, \U + \U \, j^{(1)}_{\tau}\right\} + \left(\frac{1}{{\z}} - 1\right) \cdot \left\{j^{(1)\,\st}_{\tau} \, \U + \U \, j^{(3)}_{\tau}\right\}, \qquad \label{IntegrableBCs2}
\end{IEEEeqnarray}
for the string sigma model \eqref{StringActionOSP226} in the presence of a probe D-brane at $\sigma = 0$.
\section[The D2-D4 intersection]{The D2-D4 intersection \label{Section:D2D4intersection}}
\indent The D2-probe-D4 brane system consists of $N$ coincident D2-branes intersecting a single (probe) D4-brane. The relative orientation of the branes is the following \cite{FujitaLiRyuTakayanagi09}:
\renewcommand{\arraystretch}{1.1}
\begin{center}\begin{tabular}{|c||c|c|c|c|c|c|c|c|c|c|}
\hline
& $x_0$ & $x_1$ & $x_2$ & $z$ & $\xi$ & $\theta_1$ & $\phi_1$ & $\theta_2$ & $\phi_2$ & $\psi$ \\ \hline
\text{D2} & $\bullet$ & $\bullet$ & $\bullet$ &&&&&&& \\ \hline
\text{D4} & $\bullet$ && $\bullet$ & $\bullet$ && $\bullet$ & $\bullet$ &&& \\ \hline
\end{tabular}\end{center}
Equivalently we may consider a single D4-brane inside AdS$_4 \times \CP^3$, the metric of which is given by
\begin{IEEEeqnarray}{c}
ds^2 = \frac{\ell^2}{z^2} \left(-dx_0^2 + dx_1^2 + dx_2^2 + dz^2\right) + 4\ell^2 ds^2_{\CP^3}, \label{MetricAdS4xCP3}
\end{IEEEeqnarray}
where, for $\xi \in \left[0,\pi/2\right)$, $\theta_{1,2} \in \left[0,\pi\right]$, $\phi_{1,2} \in \left[0,2\pi\right)$ and $\psi \in \left[-2\pi,2\pi\right]$,
\begin{IEEEeqnarray}{ll}
ds_{\CP^3}^2 = d\xi^2 &+ \cos^2\xi \, \sin^2\xi \left(d\psi + \frac{1}{2} \cos\theta_1 \, d\phi_1 - \frac{1}{2} \cos\theta_2 \, d\phi_2\right)^2 + \nonumber \\
&+ \frac{1}{4} \cos^2\xi \Big(d\theta_1^2 + \sin^2 \theta_1 \, d\phi_1^2\Big) + \frac{1}{4}\sin^2\xi \left(d\theta_2^2 + \sin^2\theta_2 \, d\phi_2^2\right). \qquad \label{MetricCP3}
\end{IEEEeqnarray}
In this background the probe D4-brane wraps an AdS$_3 \times \CP^1$ subset of AdS$_4 \times \CP^3$, supported by $Q$ units of magnetic flux through $\CP^1$. The presence of the flux forces exactly $q \equiv \sqrt{2\lambda}\,Q$ of the D2-branes to terminate on one side of the D4-brane. \\
\indent The embedding of the probe D4-brane in AdS$_4\times\CP^3$ is described by the set of equations \cite{ChandrasekharPanda09}
\begin{IEEEeqnarray}{c}
x_2 \eq Q \cdot z \qquad \& \qquad \xi \eq 0, \qquad \theta_2,\, \phi_2,\, \psi \eq \text{constant}, \qquad \label{D4braneEmbedding}
\end{IEEEeqnarray}
while the field strength of the worldvolume gauge field $F$ is
\begin{IEEEeqnarray}{c}
F = \ell^2 \, Q \, d\cos\theta_1 \wedge d\phi_1 = -\ell^2 \, Q \, \sin\theta_1 \, d\theta_1 \, d\phi_1 = dA, \qquad \label{MagneticField1}
\end{IEEEeqnarray}
and the corresponding gauge potential
\begin{IEEEeqnarray}{c}
A = \ell^2 Q \left(c + \cos\theta_1\right)d\phi_1, \qquad \label{MagneticPotential}
\end{IEEEeqnarray}
where $c$ is a constant. Notice that $\CP^1$ is just a 2-sphere of radius $\ell$:
\begin{IEEEeqnarray}{l}
ds^2_{\CP^1} = \ell^2\left(d\theta_1^2 + \sin^2\theta_1 \, d\phi_1^2\right) = \sum_{i = 4}^6 dx_{i} \, dx_{i}, \qquad \sum_{i = 4}^6 x_{i} \, x_{i} = \ell^2, \qquad
\end{IEEEeqnarray}
so that, in terms of the Cartesian coordinates $x_{4,5,6}$, the field strength $F$ is written as
\begin{IEEEeqnarray}{c}
F = -\frac{Q}{\ell}\,\left(x_4 \, dx_5 \wedge dx_6 + x_5 \, dx_6 \wedge dx_4 + x_6 \, dx_4 \wedge dx_5\right). \qquad \label{MagneticField2}
\end{IEEEeqnarray}
Componentwise,
\begin{IEEEeqnarray}{c}
F_{ij} = -\frac{Q}{\ell}\,\epsilon_{ijk}x_k = \partial_i A_j - \partial_j A_i, \qquad \label{MagneticField3}
\end{IEEEeqnarray}
where $A_i$ are the $i = 4,5,6$ components of the 1-form gauge potential \eqref{MagneticPotential}.
\paragraph{Coset parametrization} In what follows we only consider bosonic strings in AdS$_4 \times \CP^3$. The coset representative of AdS$_4 \times \CP^3$ is taken to be \cite{DekelOz11b}:
\begin{IEEEeqnarray}{c}
\mathfrak{g} = \left(\begin{array}{cc} e^{P_{\mu}x^{\mu}}z^{D} & 0 \\ 0 & e^{-R_8 \psi} e^{T_3 \phi_1} e^{T_4 \left(\theta_1 + \frac{\pi}{2}\right)} e^{R_3 \phi_2} e^{R_4 \left(\theta_2 + \frac{\pi}{2}\right)} e^{2T_6 \xi} \end{array}\right), \qquad \label{CosetRepresentativeAdS4xCP3}
\end{IEEEeqnarray}
where $D$ and $P_{\mu}$ ($\mu = 0,1,2$) are the conformal generators of dilations and translations respectively. $R_1,\ldots,R_9$ are the graded-0 generators of $\mathfrak{so}\left(6\right)$ with respect to its $\mathfrak{u}\left(3\right)$ subalgebra, and $T_1,\ldots,T_6$ are the graded-2 generators. All the relevant conventions can be found in appendices \ref{Appendix:GammaMatrices}--\ref{Appendix:T-Matrices}. The $J^{(2)}$ components of the moving-frame current $J = \mathfrak{g}^{-1}d\mathfrak{g}$ are given by:
\begin{IEEEeqnarray}{ll}
J^{(2)}_{\text{AdS}} = & \frac{1}{2z}\left(2 D dz + \left(P_{\mu} + K_{\mu}\right)dx^{\mu}\right) \label{MovingCurrentsAdS4xCP3a} \\[6pt]
J^{(2)}_{\CP} = & T_1 d\theta_2 \sin\xi + T_2 d\phi_2 \sin\theta_2 \sin\xi - T_3 d\phi_1 \sin\theta_1 \cos\xi + T_4 d\theta_1 \cos\xi - \nonumber \\[6pt]
& - T_5 \sin\xi\cos\xi\left(2 d\psi + d\phi_1 \cos\theta_1 - d\phi_2 \cos\theta_2\right) + 2 T_6 d\xi, \label{MovingCurrentsAdS4xCP3b}
\end{IEEEeqnarray}
where $K_{\mu}$ are the generators of special conformal transformations (cf.\ appendix \ref{Appendix:GammaMatrices}). As a crosscheck of the coset parametrization \eqref{CosetRepresentativeAdS4xCP3}, we find that the supertrace $\ell^2\,\text{tr}\Big[\big(J^{(2)}_{\text{AdS}}\big)^2 - \big(J^{(2)}_{\CP}\big)^2\Big]$ correctly reproduces the AdS$_4 \times \CP^3$ metric \eqref{MetricAdS4xCP3}--\eqref{MetricCP3} as it should.
\paragraph{D2-D4 boundary conditions} With the coset parametrization \eqref{CosetRepresentativeAdS4xCP3} and the fermions switched off, the action \eqref{StringActionOSP226} is just the string Polyakov action:
\begin{IEEEeqnarray}{c}
S = -\frac{1}{4\pi\alpha'}\int d\tau d\sigma G_{mn} \gamma^{\alpha\beta} \partial_{\alpha} X^{m} \partial_{\beta} X^{n} + \int d\tau \left[A_i\dot{X}_i\right]_{\sigma=0}, \qquad \label{StringPolyakovAction}
\end{IEEEeqnarray}
where $G_{mn}$ is the AdS$_4 \times \CP^3$ metric tensor \eqref{MetricAdS4xCP3}--\eqref{MetricCP3} and $\gamma_{\alpha\beta} = \text{diag}\left(-1,1\right)$ in the conformal gauge. The boundary term at $\sigma = 0$ couples the string to the Maxwell field of the D4-brane \eqref{MagneticField1}--\eqref{MagneticPotential}. \\
\indent By varying the action \eqref{StringPolyakovAction} it can be shown that the boundary conditions (BCs) are Dirichlet for the string coordinates ($X_a$) that are transverse to the D4-brane and mixed Neumann-Dirichlet for the longitudinal coordinates ($X_i$) of the string:
\begin{IEEEeqnarray}{rl}
\acute{X}_{i} - 2\pi\alpha' F_{ij} \dot{X}_{j} \eq 0 \qquad &\text{(longitudinal: Neumann-Dirichlet)} \label{String_BCs1} \\
\dot{X}_a \eq 0 \qquad &\text{(transverse: Dirichlet)}. \label{String_BCs2}
\end{IEEEeqnarray}
For the AdS$_4$ coordinates of the string ($x_{0,1,2},\, z$), \eqref{String_BCs1}-\eqref{String_BCs2} give
\begin{IEEEeqnarray}{rl}
\acute{x}_{0,1} \eq \acute{z} + Q \, \acute{x}_2 \eq 0 \qquad & \text{(Neumann)} \label{BoundaryConditionsAdS1} \\
\dot{x}_2 - Q \, \dot{z} \eq 0 \qquad & \text{(Dirichlet)}, \label{BoundaryConditionsAdS2}
\end{IEEEeqnarray}
while for the compact coordinates $x_{i}$ ($i = 4,5,6$) we obtain
\begin{IEEEeqnarray}{c}
\ell\acute{x}_i \eq \tilde{Q} \, \epsilon_{ijk}x_j\dot{x}_k, \qquad \tilde{Q} \equiv \frac{q}{\lambda}. \qquad \label{BoundaryConditionsCP1}
\end{IEEEeqnarray}
The BCs on $\CP^3$ can then be expressed in terms of the coordinates \eqref{MetricCP3} as follows:
\begin{IEEEeqnarray}{rl}
\acute{\theta}_1 + \tilde{Q}\sin\theta_1\dot{\phi}_1 \eq \sin\theta_1\acute{\phi}_1 - \tilde{Q} \, \dot{\theta}_1 \eq 0 \qquad & \text{(Neumann-Dirichlet)} \qquad \label{BoundaryConditionsCP2} \\
\dot{\xi} \eq \dot{\theta}_2 \eq \dot{\phi}_2 \eq \dot{\psi} \eq0 \qquad & \text{(Dirichlet)}. \qquad \label{BoundaryConditionsCP3}
\end{IEEEeqnarray}
Apart from the BCs \eqref{BoundaryConditionsAdS1}--\eqref{BoundaryConditionsAdS2} and \eqref{BoundaryConditionsCP2}--\eqref{BoundaryConditionsCP3}, the string coordinates on the D4-brane should also obey \eqref{D4braneEmbedding}.
\section[D4-brane integrability]{D4-brane integrability \label{Section:D4braneIntegrability}}
To show that the boundary conditions \eqref{BoundaryConditionsAdS1}--\eqref{BoundaryConditionsAdS2} and \eqref{BoundaryConditionsCP2}--\eqref{BoundaryConditionsCP3} are classically integrable, a dynamical reflection matrix $\U\left(\tau;\x\right)$ must be specified that satisfies the bosonic part of the integrability condition \eqref{IntegrableBCs2}:
\begin{IEEEeqnarray}{c}
\dot{\U} \eq \frac{2}{\x^2 - 1} \cdot \left\{j^{(2)\,\text{t}}_{\tau} \, \U + \U \, j^{(2)}_{\tau}\right\} + \frac{2\x}{\x^2 - 1} \cdot \left\{j^{(2)\,\text{t}}_{\sigma} \, \U - \U \, j^{(2)}_{\sigma}\right\}, \qquad \label{IntegrableBCs3}
\end{IEEEeqnarray}
upon imposing the BCs. The reflection matrix $\U$ should be block diagonal, with the upper block corresponding to the AdS$_4$ space, and the lower block corresponding to $\CP^3$:
\begin{IEEEeqnarray}{c}
\U = \left[\begin{array}{cc} \U_{\text{AdS}} & 0 \\ 0 & \U_{\CP} \end{array}\right].
\end{IEEEeqnarray}
\indent The boundary values of the bosonic $\mathds{Z}_4$ components of the fixed-frame current $j^{(2)} = \mathfrak{g} J^{(2)} \mathfrak{g}^{-1}$ that show up in the integrability condition \eqref{IntegrableBCs3} are, in the case of AdS$_4$,
\small\begin{IEEEeqnarray}{ll}
j^{(2)}_\tau =& \frac{1}{2z^2}\left[2\left(z\dot{z} + x^{\mu}\dot{x}_{\mu}\right) (D - x^{\nu} P_{\nu}) + \left(z^2 + x^2\right) \dot{x}^{\mu} P_{\mu} + \dot{x}^{\mu} K_{\mu} + x^\mu \dot{x}^\nu L_{\mu\nu}\right] \qquad \label{FixedCurrentAdStau} \\[6pt]
j^{(2)}_\sigma \eq& \frac{\acute{x}_2}{2z^2} \left((z^2 + x^2) P_2 + K_2 + x^\mu L_{\mu2}\right), \qquad \label{FixedCurrentAdSsigma}
\end{IEEEeqnarray}\normalsize
where $L_{\mu\nu}$ are the conformal generators of rotations. Note that the $j^{(2)}_{\tau}$ component of the fixed-frame current has the same form in the bulk and the boundary. For $\CP^3$, we find
\small\begin{IEEEeqnarray}{ll}
j^{(2)}_\tau \eq& T_3 \dot{\phi}_1 \sin^2\theta_1 + T_4 \left(\dot{\theta}_1\cos\phi_1 - \dot{\phi}_1\sin\theta_1\cos\theta_1\sin\phi_1\right) - \nonumber \\
& \hspace{4cm} - \frac{1}{4}\left(R_7 + R_9\right)\left(2\dot{\theta}_1\sin\phi_1 + \dot{\phi}_1\sin2\theta_1\cos\phi_1\right) \qquad \label{FixedCurrentCPtau} \\[6pt]
j^{(2)}_\sigma \eq& -\tilde{Q}\cdot\frac{d}{d\tau}\left[T_3 \cos\theta_1 + T_4 \sin\theta_1\sin\phi_1 + \frac{1}{2}\left(R_7 + R_9\right) \sin\theta_1\cos\phi_1 - R_8\right] + \nonumber \\[6pt]
& + f\left(\theta_1,\theta_2,\phi_1,\phi_2,\psi\right)\cdot \xi', \label{FixedCurrentCPsigma} \qquad
\end{IEEEeqnarray}\normalsize
where we have defined,
\small\begin{IEEEeqnarray}{ll}
f\left(\theta_1,\theta_2,\phi_1,\phi_2,\psi\right) &= T_1\left(a_-\cos\psi - b_+\sin\psi\right) + T_2\left(a_-\sin\psi + b_+\cos\psi\right) - \nonumber \\
& - T_5\left(c_+\cos\psi + d_-\sin\psi\right) - T_6\left(c_+\sin\psi - d_-\cos\psi\right) - \nonumber \\
& - R_1\left(a_+\cos\psi - b_-\sin\psi\right) + R_2\left(a_+\sin\psi + b_-\cos\psi\right) + \nonumber \\
& + R_5\left(c_-\cos\psi + d_+\sin\psi\right) + R_6\left(c_-\sin\psi - d_+\cos\psi\right) \qquad \label{CoefficientXiPrime1}
\end{IEEEeqnarray}\normalsize
and for $\theta_{\pm} = \left(\theta_1 \pm \theta_2\right)/2$, $\phi_{\pm} = \left(\phi_1 \pm \phi_2\right)/2$,
\small\begin{IEEEeqnarray}{ll}
a_{\pm} = \cos\theta_+\cos\phi_- \pm \sin\theta_-\cos\phi_+, \quad &b_{\pm} = \sin\theta_+\sin\phi_+ \pm \cos\theta_-\sin\phi_- \qquad \label{CoefficientXiPrime2} \\
c_{\pm} = \sin\theta_-\sin\phi_+ \pm \cos\theta_+\sin\phi_-, \quad &d_{\pm} = \cos\theta_-\cos\phi_- \pm \sin\theta_+\cos\phi_+. \qquad \label{CoefficientXiPrime3}
\end{IEEEeqnarray}\normalsize
\subsection[Integrable boundary conditions on AdS$_4$]{Integrable boundary conditions on AdS$_4$}
As expected, the AdS$_4$ reflection matrix of the D4-brane is very similar to the one that was found in \cite{LinardopoulosZarembo21} for the AdS$_5$ component of the D5-brane:
\begin{IEEEeqnarray}{c}
\U_{\text{AdS}} = K_4 \cdot \Bigg[\gamma_2 + \frac{2Q}{\x^2 + 1}\cdot\frac{x^{\mu}\gamma_{\mu} - \Pi_{+} - \left(z^2 + x^2\right)\Pi_{-}}{z}\Bigg], \label{ReflectionMatrixAdS}
\end{IEEEeqnarray}
where the projectors $\Pi_{\pm}$ have been defined in \eqref{ProjectorsAdS} of appendix \ref{Appendix:GammaMatrices}. \\
\indent Further details about the construction of \eqref{ReflectionMatrixAdS} can be found in \cite{LinardopoulosZarembo21}. The reflection matrix \eqref{ReflectionMatrixAdS} satisfies the integrability conditions \eqref{IntegrableBCs3} upon imposing the BCs \eqref{D4braneEmbedding}, \eqref{BoundaryConditionsAdS1}--\eqref{BoundaryConditionsAdS2}, thereby proving that the AdS$_3 \subset \text{AdS}_4$ part of the probe D4-brane is integrable. In the zero-flux case ($Q = 0$) the reflection matrix is just a constant matrix, independent from the spectral parameter $\x$ and non-dynamical, as expected from the work of Dekel and Oz \cite{DekelOz11b}.
\subsection[Integrable boundary conditions on $\CP^3$]{Integrable boundary conditions on $\CP^3$}
\noindent Let us now specify the $\CP^3$ part of the reflection matrix for the probe D4-brane. Notice that the T and R-matrices that take part in the construction of the $\CP^3$ coset element and currents are all antisymmetric (cf.\ appendix \ref{Appendix:T-Matrices}). Because of this, the integrability condition \eqref{IntegrableBCs3} takes the following form in $\CP^3$:
\begin{IEEEeqnarray}{c}
\dot{\U} \eq \frac{2}{\x^2 - 1} \left(\big[\U , j^{(2)}_{\tau}\big] - \x \cdot \big\{\U , j^{(2)}_{\sigma}\big\}\right). \qquad \label{IntegrableBCs4}
\end{IEEEeqnarray}
\paragraph{Zero-flux case} In the zero-flux case, we already know from Dekel-Oz \cite{DekelOz11b} that the corresponding reflection matrix is again constant, independent from the spectral parameter $\x$ and non-dynamical. In this case the condition \eqref{IntegrableBCs4} becomes
\begin{IEEEeqnarray}{c}
\big[\U_0 , j^{(2)}_{\tau}\big] \eq \big\{\U_0 , j^{(2)}_{\sigma}\big\} \eq 0, \qquad Q = 0. \label{ReflectionEquation}
\end{IEEEeqnarray}
To solve the above equation we express $\U_0$ as a linear combination of the antisymmetric matrices T, R and their symmetric products. Solving the resulting system, we are led to the reflection matrix\footnote{The most general solution of \eqref{ReflectionEquation} contains the terms:
\begin{IEEEeqnarray}{ll}
\U_0 &= c_1\big(R_3 + R_4\tan\theta_2\sin\phi_2 + \frac{\left(R_7 - R_9\right)}{2}\tan\theta_2\cos\phi_2 + \frac{R_8}{2}\sec\theta_2\big) + \nonumber \\& + c_2\left(T_1^2 - 3T_3^2 + T_5^2\right), \qquad
\end{IEEEeqnarray}
where $c_1$, $c_2$ are constants. Demanding $\U_0$ to be involutive and symmetric gives $c_1 = 0$ and $c_2 = 2$, which leads to \eqref{ReflectionMatrixCP0}. This choice also ensures that the conserved charges will have the right amount of symmetry and supersymmetry (see section \ref{Section:ConservedCharges} below).}
\begin{IEEEeqnarray}{ll}
\U_0 = 2\left(T_1^2 - 3T_3^2 + T_5^2\right). \qquad \label{ReflectionMatrixCP0}
\end{IEEEeqnarray}
This matrix is in full agreement with the result of Dekel and Oz \cite{DekelOz11b}. It is obviously non-dynamical ($\dot{\U}_0 = 0$), symmetric ($\U_0^{\text{t}} = \U_0$), involutory ($\U_0^2 = 1$) and commutes with the $\CP^3$ coset element $[g_{\CP},\U_0] \eq 0$ on the brane. This means that the moving and the fixed-frame currents satisfy the same set of equations \eqref{ReflectionEquation}, see \cite{DekelOz11b} for more.
\paragraph{Nonzero-flux case} To determine the reflection matrix in the nonzero-flux ($Q \neq 0$) case, we first rewrite the fixed frame current \eqref{FixedCurrentCPsigma} as
\begin{IEEEeqnarray}{c}
j^{(2)}_\sigma \eq -\frac{\tilde{Q}}{2}\cdot\dot{S} + f\left(\theta_1,\theta_2,\phi_1,\phi_2,\psi\right)\cdot \xi',
\end{IEEEeqnarray}
where the function $f\left(\theta_1,\theta_2,\phi_1,\phi_2,\psi\right)$ was defined in \eqref{CoefficientXiPrime1}--\eqref{CoefficientXiPrime3} and
\begin{IEEEeqnarray}{c}
S \equiv 2T_3 \cos\theta_1 + 2T_4 \sin\theta_1\sin\phi_1 + \left(R_7 + R_9\right) \sin\theta_1\cos\phi_1 - R_8. \qquad
\end{IEEEeqnarray}
The function $S$ has some interesting properties on the brane
\begin{IEEEeqnarray}{c}
\big\{S, j^{(2)}_{\sigma}\big\} \eq \big[\U_0, j^{(2)}_{\tau}\big] \eq 0, \qquad \tilde{Q}\big[S, j^{(2)}_{\tau}\big] \eq \big\{\U_0, j^{(2)}_{\sigma}\big\} \eq - \tilde{Q}\cdot\dot{S}, \qquad \label{Sproperties1}
\end{IEEEeqnarray}
very reminiscent of the way the reflection matrix \eqref{ReflectionMatrixAdS} was constructed in \cite{LinardopoulosZarembo21}. Of course, \eqref{ReflectionEquation} is partially violated for $Q \neq 0$. We remark in passing that the anticommutator and the commutator of $S$ with the zero-flux reflection matrix $\U_0$ are given by (derivatives follow easily)
\begin{IEEEeqnarray}{c}
\big\{\U_0, S\big\} = 2S + 4R_8, \qquad \big[U_0, S\big] = 0. \label{Sproperties2}
\end{IEEEeqnarray}
Following \cite{LinardopoulosZarembo21}, we make the next ansatz for the reflection matrix in $\CP^3$:
\begin{IEEEeqnarray}{c}
\U_{\CP} = \U_0 + C S, \label{ReflectionMatrixAnsatzCP}
\end{IEEEeqnarray}
where $C$ is some function of the flux number $Q$ and the spectral parameter $\x$. By using the properties \eqref{Sproperties1} of the function $S$ and the zero-flux reflection matrix $\U_0$, we arrive at the reflection matrix of the D4-brane in $\CP^3$:
\begin{IEEEeqnarray}{c}
\U_{\CP} = \U_0 + \frac{2\tilde{Q}\,\x}{\x^2 + 1}\cdot S. \label{ReflectionMatrixCP}
\end{IEEEeqnarray}
One can readily check that \eqref{ReflectionMatrixCP} satisfies the integrability conditions \eqref{IntegrableBCs3} and \eqref{IntegrableBCs4} upon imposing the string BCs \eqref{D4braneEmbedding}, \eqref{BoundaryConditionsCP2}--\eqref{BoundaryConditionsCP3}, so that the $\CP^1 \subset \CP^3$ part of the probe D4-brane is again integrable.
\section[Conserved charges]{Conserved charges \label{Section:ConservedCharges}}
As we have already mentioned, the conserved charges of integrable field theories are encoded in their (single-row) monodromy matrices \eqref{SingleRowMonodromyMatrix}. The double-row monodromy matrix \eqref{DoubleRowMonodromyMatrix1} takes into account boundary conditions that may be imposed on the bulk fields. Taylor-expanding either monodromy matrix in the spectral parameter $\x$ generates an integrable hierarchy of conserved charges. The way the double-row monodromy matrix is built out of two monodromy matrices and the reflection matrix \eqref{DoubleRowMonodromyMatrix1} suggests that the double-row charge hierarchy will generally be a subset of the single-row charge hierarchy.
\paragraph{Global symmetries} The conserved charges of the string sigma model \eqref{StringActionOSP226} can be obtained by expanding the monodromy matrix \eqref{SingleRowMonodromyMatrix}. We first note that \eqref{SingleRowMonodromyMatrix} can be written as (omitting all $\tau$-dependencies for simplicity):
\begin{IEEEeqnarray}{c}
\M\left(\x\right) = \mathfrak{g}\left(0\right) \cdot \Pexp\left(\int_{0}^{\infty} ds \, L_{\sigma}\left(s ; \x\right)\right) = \Pexp\left(\int_{0}^{\infty} ds \, a_{\sigma}\left(s ; \x\right)\right). \qquad
\end{IEEEeqnarray}
Taylor-expanding the path-ordered exponential around $\x = \infty$ leads to
\small\begin{IEEEeqnarray}{ll}
\Pexp\left(\int_{0}^{\infty} ds \, a_{\sigma}\right) &= \mathbbm{1} - \frac{2}{\x}\int_0^{\infty} ds \left[j^{(2)}_{\tau} + \frac{j^{(3)}_{\sigma} - j^{(1)}_{\sigma}}{2}\right] + \frac{2}{\x^2} \Bigg\{\int_0^{\infty} ds \bigg[j^{(2)}_{\sigma} + \nonumber \\
&\hspace{.4cm} + \frac{j^{(1)}_{\sigma} + j^{(3)}_{\sigma}}{4}\bigg] + \int_0^{\infty}\int_0^{s} ds ds' \left[2j^{(2)'}_{\tau} j^{(2)}_{\tau} + \ldots\right]\Bigg\} - \ldots \nonumber \\[6pt]
& \hspace{-1.5cm}\equiv \exp\left[2\sum_{r = 0}^{\infty} \left(-\frac{1}{\x}\right)^{r+1} \Q_{r}\right] = \mathbbm{1} - \frac{2}{\x}\,\Q_{0} + \frac{2}{\x^2}\left(\Q_{1} + \Q_{0}^2\right) - \ldots \label{SingleRowMonodromyMatrixExpansion}
\end{IEEEeqnarray}\normalsize
In the absence of fermions, the first charge in the above hierarchy is just the Noether charge of the global bosonic symmetry $SO\left(3,2\right) \times SO(6)$:
\begin{IEEEeqnarray}{c}
\Q_{0} = \int_0^{\infty} ds j_{\tau}^{(2)}, \label{GlobalCharge0}
\end{IEEEeqnarray}
where $j_{\tau}^{(2)}$ is given by \eqref{FixedCurrentAdStau} in the case of AdS. The corresponding expression for the charge $\Q_0$ will obviously involve all the $\mathfrak{so}\left(3,2\right)$ generators \eqref{GeneratorsAdS} with coefficients that are functions of the AdS coordinates and their $\tau$-derivatives. The expression of the fixed-frame current in $\CP^3$ is too complicated to be written out here. However it suffices to say that again the current (and therefore the charge $\Q_0$) involves all the $\mathfrak{so}\left(6\right)$ generators $R_{1,\ldots,9}$ and $T_{1,\ldots,6}$, with nonzero coefficients that depend on the $\CP^3$ coordinates and their $\tau$-derivatives.
\paragraph{Broken symmetries} The conserved charges of the D4-brane can be obtained by expanding the double-row monodromy matrix \eqref{DoubleRowMonodromyMatrix1}. The presence of the reflection matrix $\U$ amid two monodromy matrices $\M$ causes some of the sigma model charges (encoded in $\M$) to get cancelled by folding. To expand the double-row monodromy matrix,
\begin{IEEEeqnarray}{c}
\T\left(\x\right) = \expP \left(\int_{0}^{\infty }ds\,a_\sigma^{\st}(s;-\x)\right) \cdot \U(\x) \cdot \Pexp \left(\int_{0}^{\infty} ds\,a_\sigma(s;\x)\right), \qquad \label{DoubleRowMonodromyMatrix2}
\end{IEEEeqnarray}
in $1/\x$, we notice that the AdS$_4\times\CP^3$ reflection matrices \eqref{ReflectionMatrixAdS}, \eqref{ReflectionMatrixCP} have the following general form:
\begin{IEEEeqnarray}{c}
\U\left(\x\right) = \U_0 + \frac{1}{\x^2 + 1}\left(\x \, \U_1 + \U_2\right). \label{ReflectionMatrixGeneric}
\end{IEEEeqnarray}
Inserting \eqref{ReflectionMatrixGeneric} and the Taylor expansion \eqref{SingleRowMonodromyMatrixExpansion} of the single-row monodromy matrix $\M$ into \eqref{DoubleRowMonodromyMatrix2}, we obtain the Taylor expansion of the double-row monodromy matrix around $\x = \infty$:
\begin{IEEEeqnarray}{ll}
T\left(\x\right) &= \U_0 + \frac{1}{\x}\left\{\U_1 + \int_0^{\infty}ds \left[2\big\langle j^{(2)}_{\tau}, \U_0\big\rangle_- + \big\langle \big(j^{(3)}_{\sigma} - j^{(1)}_{\sigma}\big), \U_0\big\rangle_-\right]\right\} + \ldots \nonumber \\[6pt]
& \equiv \U_0 + \frac{2}{\x}\tilde{\Q}_{0} + \frac{2}{\x^2} \left(\tilde{\Q}_{1} + \tilde{\Q}_{0}^2\right) + \ldots,
\end{IEEEeqnarray}
where the supertransposition bracket $\left\langle\,\right\rangle_-$ has been defined as
\begin{IEEEeqnarray}{c}
\left\langle A, B\right\rangle_- \equiv A^{\st} B - B A.
\end{IEEEeqnarray}
The subset of the $OSP\left(2,2|6\right)$ global symmetry that is preserved by the D4-brane is determined by the set of bosonic and supersymmetric charges that span the first conserved charge $\tilde{\Q}_0$:
\begin{IEEEeqnarray}{c}
\tilde{\Q}_{0} = \frac{\U_1}{2} + \int_0^{\infty}ds \,\left[\big\langle j^{(2)}_{\tau}, \U_0\big\rangle_- + \frac{1}{2}\big\langle \big(j^{(3)}_{\sigma} - j^{(1)}_{\sigma}\big), \U_0\big\rangle_-\right]. \qquad \label{GlobalCharge1}
\end{IEEEeqnarray}
Broken symmetries get cancelled by folding and are absent from either $\U_1$ or the supertransposition brackets $\big\langle j^{(1,2,3)}, \U_0\big\rangle_-$. \\
\indent To determine the preserved bosonic symmetries by the D4-brane in AdS$_4$, we need to remove the fermionic currents from \eqref{GlobalCharge1}, replace supertransposition with simple transposition and read off $\U_0 = \gamma_1$ and $\U_1 = 0$ from \eqref{ReflectionMatrixAdS}. Noting that the fixed-frame current \eqref{FixedCurrentAdStau} is made up from all the conformal generators \eqref{GeneratorsAdS}, the global charges of $\mathfrak{so}\left(3,2\right)$ that get cancelled are
\begin{IEEEeqnarray}{c}
\left\langle P_2, \U_0\right\rangle_- = \left\langle K_2, \U_0\right\rangle_- = \left\langle L_{i2}, \U_0\right\rangle_- = 0,
\end{IEEEeqnarray}
where $i = 0,1$. The unbroken conformal generators are thus
\begin{IEEEeqnarray}{c}
\big\{D,P_i,K_i,L_{01}\big\},
\end{IEEEeqnarray}
and span the subgroup $SO(2,2) \subset SO(3,2)$, a result which is consistent with the AdS$_3$ geometry of the D4-brane in AdS$_4$. In $\CP^3$ the $\mathfrak{so}\left(6\right)$ generators R and T that make up the fixed-frame current $j^{(2)}_{\tau}$ are all antisymmetric so that the supertransposition bracket in \eqref{GlobalCharge1} can be replaced by an anticommutator:
\begin{IEEEeqnarray}{c}
\tilde{\Q}_{0} = \frac{\U_1}{2} - \int_0^{\infty}ds \left\{ j^{(2)}_{\tau}, \U_0\right\}.
\end{IEEEeqnarray}
Moreover, \eqref{ReflectionMatrixCP0} and \eqref{ReflectionMatrixCP} imply $\U_0 = 2\left(T_1^2 - 3T_3^2 + T_5^2\right)$ and $\U_1 = 2\tilde{Q} S$ so
\begin{IEEEeqnarray}{c}
\left\{ T_{1,2,5,6}, \U_0\right\} = \left\{ R_{1,2,5,6}, \U_0\right\} = 0,
\end{IEEEeqnarray}
which is again consistent with the $\CP^1$ geometry of the D4-brane in $\CP^3$. Indeed, it is quite straightforward to show that the sets of unbroken generators
\begin{IEEEeqnarray}{c}
\big\{T_3,T_4,-\frac{R_7 + R_9}{2}\big\}\times\big\{R_3,R_4,-\frac{R_7 - R_9}{2}\big\}\times\big\{R_8\big\},
\end{IEEEeqnarray}
span the subgroup $SU(2)\times SU(2)\times U(1) \subset SO(6)$.
\paragraph{Supersymmetries} To identify the conserved supercharges, we note that the fermionic currents of $\mathfrak{osp}\left(2,2|6\right)$ have the form
\begin{IEEEeqnarray}{c}
j = \left[\begin{array}{cc} 0 & \mathcal{Q} \\ -\mathcal{Q}^{\text{t}}K_4 & 0 \end{array}\right], \qquad j^{\st} = \left[\begin{array}{cc} 0 & K_4^{\text{t}}\mathcal{Q} \\ \mathcal{Q}^{\text{t}} & 0 \end{array}\right] \label{Supercurrent}
\end{IEEEeqnarray}
and consist of a total of 24 supercharges since the $4\times 6$ matrix $\mathcal{Q}$ also obeys the reality condition
\begin{IEEEeqnarray}{c}
\mathcal{Q}^* = i\gamma_3 \mathcal{Q}.
\end{IEEEeqnarray}
The broken supercharges are once more determined from
\begin{IEEEeqnarray}{c}
\big\langle j, \U_0\big\rangle_- = 0, \qquad \label{BrokenSupercharges}
\end{IEEEeqnarray}
where $\U_0$ is the zero-flux reflection matrix that is given by \eqref{ReflectionMatrixAdS}, \eqref{ReflectionMatrixCP0}:
\begin{IEEEeqnarray}{c}
\U_0 = \left[\begin{array}{cc} \gamma_1 & 0 \\ 0 & 2\left(T_1^2 - 3T_3^2 + T_5^2\right) \end{array}\right]. \label{ReflectionMatrixAdS4xCP3}
\end{IEEEeqnarray}
In \cite{DekelOz11b} it was shown that the zero-flux reflection matrix \eqref{ReflectionMatrixAdS4xCP3} is consistent with the inclusion of fermionic degrees of freedom in the coset element \eqref{CosetRepresentativeAdS4xCP3}. Inserting \eqref{Supercurrent} and \eqref{ReflectionMatrixAdS4xCP3} into \eqref{BrokenSupercharges}, we are led to
\begin{IEEEeqnarray}{c}
2K_4^{\text{t}}\,\mathcal{Q} \left(T_1^2 - 3T_3^2 + T_5^2\right) = \gamma_1 \, \mathcal{Q},
\end{IEEEeqnarray}
which leaves us with 12 independent supercharges, exactly as it should for a half-BPS brane \cite{ChandrasekharPanda09}.
\section[Conclusions]{Conclusions \label{Section:Conclusions}}
The integrability of the string theory dual of the ABJM domain wall is a considerable indication of this system's integrability from weak to strong coupling, for all values of the bond dimension $q$. It would be interesting if this result could be supported by further evidence from the gauge theory side. This evidence would include the extension of the closed-form overlap formulas beyond the lowest bond dimensions $q=1,2$ at tree level as well as higher loop orders. The study of defect observables with the method of supersymmetric localization that was put forward in \cite{Wang20a, KomatsuWang20}, would similarly boost the study of ABJM defects from a purely nonperturbative perspective. Among the various interesting applications of dynamical reflection matrices we could single out the study of giant gravitons in either $\mathcal{N} = 4$ SYM \cite{BissiKristjansenYoungZoubos11, JiangKomatsuVescovi19a, JiangKomatsuVescovi19b} or the ABJM theory \cite{HiranoKristjansenYoung12, YangJiangKomatsuWu21a, YangJiangKomatsuWu21b}.
\subsection*{Acknowledgements}
I'm thankful to Z.\ Bajnok, C.\ Kristjansen and K.\ Zarembo for useful discussions. I would also like to thank M.\ Axenides for comments on the manuscript.
\appendix\section[Gamma matrices]{Gamma matrices \label{Appendix:GammaMatrices}}
\noindent Consider the following 4-dimensional gamma matrices (in 4 spacetime dimensions):
\small\begin{IEEEeqnarray}{ll}
\gamma_0 = i \, \sigma_3 \otimes I_2 = \left(\begin{array}{cccc}
 i & 0 & 0 & 0 \\
 0 & i & 0 & 0 \\
 0 & 0 & -i & 0 \\
 0 & 0 & 0 & -i \\\end{array}\right), \ &
\gamma_1 = \sigma_2 \otimes \sigma_3 = \left(\begin{array}{cccc}
 0 & 0 & -i & 0 \\
 0 & 0 & 0 & i \\
 i & 0 & 0 & 0 \\
 0 & -i & 0 & 0 \\\end{array}\right) \qquad \\[6pt]
\gamma_2 = -\sigma_2 \otimes \sigma_1 = \left(\begin{array}{cccc}
 0 & 0 & 0 & i \\
 0 & 0 & i & 0 \\
 0 & -i & 0 & 0 \\
 -i & 0 & 0 & 0 \\\end{array}\right), \ &
\gamma_3 = \sigma_2 \otimes \sigma_2 = \left(\begin{array}{cccc}
 0 & 0 & 0 & -1 \\
 0 & 0 & 1 & 0 \\
 0 & 1 & 0 & 0 \\
 -1 & 0 & 0 & 0 \\\end{array}\right). \qquad
\end{IEEEeqnarray}\normalsize
These matrices satisfy the Minkowski Clifford algebra $\left\{\gamma_a,\gamma_b\right\} = 2\eta_{ab}$, where $\eta_{ab} = (-+++)$. The same metric is used for all AdS index contractions throughout the paper. We also define the matrix $K_4 \equiv \gamma_{12}$ which satisfies the identities
\begin{IEEEeqnarray}{c}
\gamma_{a}^t = K_4^{-1}\gamma_{a} K_4, \qquad \gamma_{ab}^t = K_4^{-1}\gamma_{ab} K_4 \qquad K_4^2 = -1, \qquad K_4^t = -K_4, \qquad
\end{IEEEeqnarray}
for $a,b = 0,\ldots,3$ and also
\begin{IEEEeqnarray}{c}
K_4\gamma_2 = \gamma_1, \qquad K_4\gamma_1 = -\gamma_2, \qquad \gamma_2 K_4 = -\gamma_1, \qquad \gamma_1 K_4 = \gamma_2. \qquad
\end{IEEEeqnarray}
\paragraph{Bosonic generators} The bosonic subalgebra of $\mathfrak{osp}\left(2,2|6\right)$ is $\mathfrak{sp}\left(2,2\right) \oplus \mathfrak{so}\left(6\right)$. The 10 generators of the conformal algebra $\mathfrak{so}\left(3,2\right) \sim \mathfrak{sp}\left(2,2\right)$ are
\begin{IEEEeqnarray}{c}
D \equiv \frac{\gamma_3}{2}, \quad P_{\mu} \equiv \Pi_{+}\gamma_{\mu}, \quad K_{\mu} \equiv \Pi_{-}\gamma_{\mu}, \quad L_{\mu\nu} \equiv \gamma_{\mu\nu} \equiv \frac{1}{2}\left[\gamma_{\mu},\gamma_{\nu}\right], \qquad \label{GeneratorsAdS}
\end{IEEEeqnarray}
for $\mu,\nu = 0,1,2$ and
\begin{IEEEeqnarray}{c}
\Pi_{\pm} \equiv \frac{1}{2}\left(1 \pm \gamma_3\right). \qquad \label{ProjectorsAdS}
\end{IEEEeqnarray}
\section[T and R-matrices]{T and R-matrices \label{Appendix:T-Matrices}}
\noindent The Lie algebra of $\mathfrak{so}\left(6\right)$ is generated by 15 matrices $M_{ij}$ ($i,j= 1,\ldots6$),
\begin{IEEEeqnarray}{c}
\left[M_{ij},M_{kl}\right] = \delta_{il} M_{jk} + \delta_{jk} M_{il} - \delta_{ik} M_{jl} - \delta_{jl} M_{ik},
\end{IEEEeqnarray}
where $M_{ij} \equiv E_{ij} - E_{ji}$ and $E_{ij}$ are the so-called standard-unity matrices. The $\mathfrak{u}\left(3\right)$ subalgebra of $\mathfrak{so}\left(6\right)$ is generated by the 9 antisymmetric R-matrices which are defined as
\begin{IEEEeqnarray}{lll}
R_1 = \frac{1}{2}\left(M_{13} + M_{24}\right), \quad & R_2 = \frac{1}{2}\left(M_{23} - M_{14}\right), \quad & R_3 = \frac{1}{2}\left(M_{15} + M_{26}\right) \qquad \\[6pt]
R_4 = \frac{1}{2}\left(M_{25} - M_{16}\right), \quad & R_5 = \frac{1}{2}\left(M_{35} + M_{46}\right), \quad & R_6 = \frac{1}{2}\left(M_{45} - M_{36}\right) \qquad \\[12pt]
R_7 = M_{12}, \quad & R_8 = M_{34}, \quad & R_9 = M_{56}. \qquad
\end{IEEEeqnarray}
These are the graded-0 generators of $\mathfrak{so}\left(6\right)$ with respect to $\mathfrak{u}\left(3\right)$. The graded-2 generators are the 6 antisymmetric T-matrices that belong to the orthogonal space of $\mathfrak{u}\left(3\right)$ inside $\mathfrak{so}\left(6\right)$:
\begin{IEEEeqnarray}{lll}
T_1 = \frac{1}{2}\left(M_{13} - M_{24}\right), \quad & T_2 = \frac{1}{2}\left(M_{14} + M_{23}\right), \quad & T_3 = \frac{1}{2}\left(M_{15} - M_{26}\right) \qquad \\[6pt]
T_4 = \frac{1}{2}\left(M_{16} + M_{25}\right), \quad & T_5 = \frac{1}{2}\left(M_{35} - M_{46}\right), \quad & T_6 = \frac{1}{2}\left(M_{36} + M_{45}\right). \qquad
\end{IEEEeqnarray}
The T-matrices anticommute, while the R-matrices commute with $K_6$:
\begin{IEEEeqnarray}{c}
\left[T_a, K_6\right] = \left\{R_a, K_6\right\} = 0, \qquad K_6^2 = -1.
\end{IEEEeqnarray}

\noindent
\bibliographystyle{StyleABJM}
\bibliography{BibliographyABJM}

\providecommand{\href}[2]{#2}\begingroup\raggedright\begin{thebibliography}{10}

\bibitem{ABJM08}
O.~Aharony, O.~Bergman, D.~L. Jafferis, and J.~Maldacena, {\it {$\mathcal{N} =
  6$ superconformal Chern-Simons-matter theories, M2-branes and their gravity
  duals}},  {\em JHEP} {\bf \textbf{10}} (2008) 091,
  [\href{http://arxiv.org/abs/0806.1218}{{\tt arXiv:0806.1218}}].

\bibitem{KristjansenStaudacherTseytlin09}
C.~Kristjansen, M.~Staudacher, and A.~Tseytlin, {\it {Gauge-string duality and
  integrability: Progress and outlook}},  {\em J. Phys.} {\bf A42} (2009)
  250301.

\bibitem{Beisertetal12}
N.~Beisert et~al., {\it {Review of AdS/CFT integrability: An overview}},  {\em
  Lett. Math. Phys.} {\bf \textbf{99}} (2012) 3,
  [\href{http://arxiv.org/abs/1012.3982}{{\tt arXiv:1012.3982}}].

\bibitem{Klose10}
T.~Klose, {\it {Review of AdS/CFT integrability, Chapter IV.3: $\mathcal{N} =
  6$ Chern-Simons and strings on AdS$_4\times\mathbb{CP}^3$}},  {\em Lett.
  Math. Phys.} {\bf \textbf{99}} (2012) 401,
  [\href{http://arxiv.org/abs/1012.3999}{{\tt arXiv:1012.3999}}].

\bibitem{ArutyunovFrolov08}
G.~Arutyunov and S.~Frolov, {\it {Superstrings on AdS$_4 \times \mathbb{CP}^3$
  as a coset sigma model}},  {\em JHEP} {\bf \textbf{09}} (2008) 129,
  [\href{http://arxiv.org/abs/0806.4940}{{\tt arXiv:0806.4940}}].

\bibitem{Stefanski08}
B.~{Stefa\'{n}ski Jr.}, {\it {Green-Schwarz action for type IIA strings on
  AdS$_4 \times \mathbb{CP}^3$}},  {\em Nucl. Phys.} {\bf \textbf{B808}} (2009)
  80, [\href{http://arxiv.org/abs/0806.4948}{{\tt arXiv:0806.4948}}].

\bibitem{MinahanZarembo09}
J.~A. Minahan and K.~Zarembo, {\it {The Bethe ansatz for superconformal
  Chern-Simons}},  {\em JHEP} {\bf \textbf{09}} (2008) 040,
  [\href{http://arxiv.org/abs/0806.3951}{{\tt arXiv:0806.3951}}].

\bibitem{BenaPolchinskiRoiban03}
I.~Bena, J.~Polchinski, and R.~Roiban, {\it {Hidden symmetries of the
  AdS$_5\times\text{S}^5$ superstring}},  {\em Phys. Rev.} {\bf \textbf{D69}}
  (2004) 046002, [\href{http://arxiv.org/abs/hep-th/0305116}{{\tt
  hep-th/0305116}}].

\bibitem{MinahanZarembo03}
J.~A. Minahan and K.~Zarembo, {\it {The Bethe ansatz for $\mathcal{N} = 4$
  super Yang-Mills}},  {\em JHEP} {\bf \textbf{03}} (2003) 013,
  [\href{http://arxiv.org/abs/hep-th/0212208}{{\tt hep-th/0212208}}].

\bibitem{GromovKazakovLeurentVolin13}
N.~Gromov, V.~Kazakov, S.~Leurent, and D.~Volin, {\it {Quantum spectral curve
  for planar $\mathcal{N} = 4$ super Yang-Mills theory}},  {\em Phys. Rev.
  Lett.} {\bf \textbf{112}} (2014) {011602},
  [\href{http://arxiv.org/abs/1305.1939}{{\tt arXiv:1305.1939}}].

\bibitem{CavagliaFioravantiGromovTateo14}
A.~Cavagli\`{a}, D.~Fioravanti, N.~Gromov, and R.~Tateo, {\it {Quantum spectral
  curve of the $\mathcal{N} = 6$ supersymmetric Chern-Simons theory}},  {\em
  Phys. Rev. Lett.} {\bf 113} (2014) 021601,
  [\href{http://arxiv.org/abs/1403.1859}{{\tt arXiv:1403.1859}}].

\bibitem{Zoubos10}
K.~Zoubos, {\it {Review of AdS/CFT Integrability, Chapter IV.2: Deformations,
  orbifolds and open boundaries}},  {\em Lett. Math. Phys.} {\bf \textbf{99}}
  (2012) 375, [\href{http://arxiv.org/abs/1012.3998}{{\tt arXiv:1012.3998}}].

\bibitem{KarchRandall01a}
A.~Karch and L.~Randall, {\it {Localized gravity in string theory}},  {\em
  Phys. Rev. Lett.} {\bf \textbf{87}} (2001) 061601,
  [\href{http://arxiv.org/abs/hep-th/0105108}{{\tt hep-th/0105108}}].

\bibitem{KarchRandall01b}
A.~Karch and L.~Randall, {\it {Open and closed string interpretation of susy
  CFT's on branes with boundaries}},  {\em JHEP} {\bf \textbf{06}} (2001) 063,
  [\href{http://arxiv.org/abs/hep-th/0105132}{{\tt hep-th/0105132}}].

\bibitem{DeWolfeFreedmanOoguri01}
O.~DeWolfe, D.~Freedman, and H.~Ooguri, {\it {Holography and defect conformal
  field theories}},  {\em Phys. Rev.} {\bf \textbf{D66}} (2002) 025009,
  [\href{http://arxiv.org/abs/hep-th/0111135}{{\tt hep-th/0111135}}].

\bibitem{DeWolfeMann04}
O.~DeWolfe and N.~Mann, {\it {Integrable open spin chains in defect conformal
  field theory}},  {\em JHEP} {\bf \textbf{04}} (2004) 035,
  [\href{http://arxiv.org/abs/hep-th/0401041}{{\tt hep-th/0401041}}].

\bibitem{ConstableMyersTafjord99}
N.~R. Constable, R.~C. Myers, and O.~Tafjord, {\it {The noncommutative bion
  core}},  {\em Phys. Rev.} {\bf \textbf{D61}} (2000) 106009,
  [\href{http://arxiv.org/abs/hep-th/9911136}{{\tt hep-th/9911136}}].

\bibitem{ConstableMyersTafjord01a}
N.~R. Constable, R.~C. Myers, and O.~Tafjord, {\it {Non-abelian brane
  intersections}},  {\em JHEP} {\bf \textbf{06}} (2001) 023,
  [\href{http://arxiv.org/abs/hep-th/0102080}{{\tt hep-th/0102080}}].

\bibitem{NagasakiTanidaYamaguchi11}
K.~Nagasaki, H.~Tanida, and S.~Yamaguchi, {\it {Holographic interface-particle
  potential}},  {\em JHEP} {\bf \textbf{01}} (2012) 139,
  [\href{http://arxiv.org/abs/1109.1927}{{\tt arXiv:1109.1927}}].

\bibitem{NagasakiYamaguchi12}
K.~Nagasaki and S.~Yamaguchi, {\it {Expectation values of chiral primary
  operators in holographic interface CFT}},  {\em Phys. Rev.} {\bf
  \textbf{D86}} (2012) 086004, [\href{http://arxiv.org/abs/1205.1674}{{\tt
  arXiv:1205.1674}}].

\bibitem{KristjansenSemenoffYoung12b}
C.~Kristjansen, G.~W. Semenoff, and D.~Young, {\it {Chiral primary one-point
  functions in the D3-D7 defect conformal field theory}},  {\em JHEP} {\bf
  \textbf{01}} (2013) 117, [\href{http://arxiv.org/abs/1210.7015}{{\tt
  arXiv:1210.7015}}].

\bibitem{deLeeuwKristjansenZarembo15}
M.~de~Leeuw, C.~Kristjansen, and K.~Zarembo, {\it {One-point functions in
  defect CFT and integrability}},  {\em JHEP} {\bf \textbf{08}} (2015) 098,
  [\href{http://arxiv.org/abs/1506.06958}{{\tt arXiv:1506.06958}}].

\bibitem{Buhl-MortensenLeeuwKristjansenZarembo15}
I.~Buhl-Mortensen, M.~de~Leeuw, C.~Kristjansen, and K.~Zarembo, {\it {One-point
  functions in AdS/dCFT from matrix product states}},  {\em JHEP} {\bf
  \textbf{02}} (2016) 052, [\href{http://arxiv.org/abs/1512.02532}{{\tt
  arXiv:1512.02532}}].

\bibitem{deLeeuwKristjansenMori16}
M.~de~Leeuw, C.~Kristjansen, and S.~Mori, {\it {AdS/dCFT one-point functions of
  the $SU(3)$ sector}},  {\em Phys. Lett.} {\bf \textbf{B763}} (2016) 197,
  [\href{http://arxiv.org/abs/1607.03123}{{\tt arXiv:1607.03123}}].

\bibitem{deLeeuwKristjansenLinardopoulos16}
M.~de~Leeuw, C.~Kristjansen, and G.~Linardopoulos, {\it {One-point functions of
  non-protected operators in the $SO(5)$ symmetric D3-D7 dCFT}},  {\em J.
  Phys.} {\bf \textbf{A50}} (2017) 254001,
  [\href{http://arxiv.org/abs/1612.06236}{{\tt arXiv:1612.06236}}].

\bibitem{deLeeuwKristjansenLinardopoulos18a}
M.~de~Leeuw, C.~Kristjansen, and G.~Linardopoulos, {\it {Scalar one-point
  functions and matrix product states of AdS/dCFT}},  {\em Phys. Lett.} {\bf
  \textbf{B781}} (2018) 238, [\href{http://arxiv.org/abs/1802.01598}{{\tt
  arXiv:1802.01598}}].

\bibitem{deLeeuwGomborKristjansenLinardopoulosPozsgay19}
M.~de~Leeuw, T.~Gombor, C.~Kristjansen, G.~Linardopoulos, and B.~Pozsgay, {\it
  {Spin chain overlaps and the twisted Yangian}},  {\em JHEP} {\bf \textbf{01}}
  (2020) 176, [\href{http://arxiv.org/abs/1912.09338}{{\tt arXiv:1912.09338}}].

\bibitem{KristjansenMullerZarembo20a}
C.~Kristjansen, D.~{M\"{u}ller}, and K.~Zarembo, {\it {Integrable boundary
  states in D3-D5 dCFT: beyond scalars}},  {\em JHEP} {\bf \textbf{08}} (2020)
  103, [\href{http://arxiv.org/abs/2005.01392}{{\tt arXiv:2005.01392}}].

\bibitem{KristjansenMullerZarembo20b}
C.~Kristjansen, D.~{M\"{u}ller}, and K.~Zarembo, {\it {Overlaps and fermionic
  dualities for integrable super spin chains}},  {\em JHEP} {\bf \textbf{03}}
  (2021) 100, [\href{http://arxiv.org/abs/2011.12192}{{\tt arXiv:2011.12192}}].

\bibitem{Buhl-MortensenLeeuwIpsenKristjansenWilhelm16a}
I.~Buhl-Mortensen, M.~de~Leeuw, A.~C. Ipsen, C.~Kristjansen, and M.~Wilhelm,
  {\it {One-loop one-point functions in gauge-gravity dualities with defects}},
   {\em Phys. Rev. Lett.} {\bf \textbf{117}} (2016) 231603,
  [\href{http://arxiv.org/abs/1606.01886}{{\tt arXiv:1606.01886}}].

\bibitem{Buhl-MortensenLeeuwIpsenKristjansenWilhelm16c}
I.~Buhl-Mortensen, M.~de~Leeuw, A.~C. Ipsen, C.~Kristjansen, and M.~Wilhelm,
  {\it {A quantum check of AdS/dCFT}},  {\em JHEP} {\bf \textbf{01}} (2017)
  098, [\href{http://arxiv.org/abs/1611.04603}{{\tt arXiv:1611.04603}}].

\bibitem{Buhl-MortensenLeeuwIpsenKristjansenWilhelm17a}
I.~Buhl-Mortensen, M.~de~Leeuw, A.~Ipsen, C.~Kristjansen, and M.~Wilhelm, {\it
  {Asymptotic one-point functions in gauge-string duality with defects}},  {\em
  Phys. Rev. Lett.} {\bf \textbf{119}} (2017) 261604,
  [\href{http://arxiv.org/abs/1704.07386}{{\tt arXiv:1704.07386}}].

\bibitem{GimenezGrauKristjansenVolkWilhelm19}
A.~Gimenez-Grau, C.~Kristjansen, M.~Volk, and M.~Wilhelm, {\it {A quantum
  framework for AdS/dCFT through fuzzy spherical harmonics on S$^4$}},  {\em
  JHEP} {\bf \textbf{04}} (2020) 132,
  [\href{http://arxiv.org/abs/1912.02468}{{\tt arXiv:1912.02468}}].

\bibitem{GomborBajnok20a}
T.~Gombor and Z.~Bajnok, {\it {Boundary states, overlaps, nesting and
  bootstrapping AdS/dCFT}},  {\em JHEP} {\bf \textbf{10}} (2020) 123,
  [\href{http://arxiv.org/abs/2004.11329}{{\tt arXiv:2004.11329}}].

\bibitem{GomborBajnok20b}
T.~Gombor and Z.~Bajnok, {\it {Boundary state bootstrap and asymptotic overlaps
  in AdS/dCFT}},  {\em JHEP} {\bf 03} (2021) 222,
  [\href{http://arxiv.org/abs/2006.16151}{{\tt arXiv:2006.16151}}].

\bibitem{KristjansenMullerZarembo21}
C.~Kristjansen, D.~{M\"{u}ller}, and K.~Zarembo, {\it {Duality relations for
  overlaps of integrable boundary states in AdS/dCFT}},  {\em JHEP} {\bf
  \textbf{09}} (2021) 004, [\href{http://arxiv.org/abs/2106.08116}{{\tt
  arXiv:2106.08116}}].

\bibitem{deLeeuwIpsenKristjansenVardinghusWilhelm17}
M.~de~Leeuw, A.~C. Ipsen, C.~Kristjansen, K.~E. Vardinghus, and M.~Wilhelm,
  {\it {Two-point functions in AdS/dCFT and the boundary conformal bootstrap
  equations}},  {\em JHEP} {\bf \textbf{08}} (2017) 020,
  [\href{http://arxiv.org/abs/1705.03898}{{\tt arXiv:1705.03898}}].

\bibitem{Widen17}
E.~Widen, {\it {Two-point functions of $SU(2)$-subsector and length-two
  operators in dCFT}},  {\em Phys. Lett.} {\bf \textbf{B773}} (2017) 435,
  [\href{http://arxiv.org/abs/1705.08679}{{\tt arXiv:1705.08679}}].

\bibitem{deLeeuwIpsenKristjansenWilhelm17}
M.~de~Leeuw, A.~C. Ipsen, C.~Kristjansen, and M.~Wilhelm, {\it {Introduction to
  integrability and one-point functions in $\mathcal{N} = 4$ SYM and its defect
  cousin}},  {\em Les Houches Lect.\ Notes} {\bf 106} (2019)
  [\href{http://arxiv.org/abs/1708.02525}{{\tt arXiv:1708.02525}}].

\bibitem{deLeeuw19}
M.~de~Leeuw, {\it {One-point functions in AdS/dCFT}},  {\em J. Phys.} {\bf
  \textbf{A53}} (2020) 283001, [\href{http://arxiv.org/abs/1908.03444}{{\tt
  arXiv:1908.03444}}].

\bibitem{Linardopoulos20}
G.~Linardopoulos, {\it {Solving holographic defects}},  {\em PoS} (2020) 141,
  [\href{http://arxiv.org/abs/2005.02117}{{\tt arXiv:2005.02117}}].

\bibitem{GaiottoWitten08a}
D.~Gaiotto and E.~Witten, {\it {Supersymmetric boundary conditions in
  $\mathcal{N} = 4$ super Yang-Mills theory}},  {\em J. Stat. Phys.} {\bf
  \textbf{135}} (2009) 789, [\href{http://arxiv.org/abs/0804.2902}{{\tt
  arXiv:0804.2902}}].

\bibitem{PiroliPozsgayVernier17}
L.~Piroli, B.~Pozsgay, and E.~Vernier, {\it {What is an integrable quench?}},
  {\em Nucl. Phys.} {\bf \textbf{B925}} (2017) 362,
  [\href{http://arxiv.org/abs/1709.04796}{{\tt arXiv:1709.04796}}].

\bibitem{GhoshalZamolodchikov93}
S.~Ghoshal and A.~B. Zamolodchikov, {\it {Boundary S-matrix and boundary state
  in two-dimensional integrable quantum field theory}},  {\em Int. J. Mod.
  Phys.} {\bf \textbf{A9}} (1994) 3841,
  [\href{http://arxiv.org/abs/hep-th/9306002}{{\tt hep-th/9306002}}]. [Erratum:
  Int. J. Mod. Phys. \textbf{A9} (1994) 4353].

\bibitem{LinardopoulosZarembo21}
G.~Linardopoulos and K.~Zarembo, {\it {String integrability of defect CFT and
  dynamical reflection matrices}},  {\em JHEP} {\bf \textbf{05}} (2021) 203,
  [\href{http://arxiv.org/abs/2102.12381}{{\tt arXiv:2102.12381}}].

\bibitem{DekelOz11b}
A.~Dekel and Y.~Oz, {\it {Integrability of Green-Schwarz sigma models with
  boundaries}},  {\em JHEP} {\bf \textbf{08}} (2011) 004,
  [\href{http://arxiv.org/abs/1106.3446}{{\tt arXiv:1106.3446}}].

\bibitem{KristjansenVuZarembo21}
C.~Kristjansen, D.-L. Vu, and K.~Zarembo, {\it {Integrable domain walls in ABJM
  theory}},  {\em JHEP} {\bf \textbf{02}} (2022) 070,
  [\href{http://arxiv.org/abs/2112.10438}{{\tt arXiv:2112.10438}}].

\bibitem{Gombor21}
T.~Gombor, {\it {On exact overlaps for $\mathfrak{gl}(N)$ symmetric spin
  chains}},  \href{http://arxiv.org/abs/2110.07960}{{\tt arXiv:2110.07960}}.

\bibitem{Sklyanin87a}
E.~K. Sklyanin, {\it {Boundary conditions for integrable equations}},  {\em
  Funct. Anal. Appl.} {\bf 21} (1987) 164.

\bibitem{ChandrasekharPanda09}
B.~Chandrasekhar and B.~Panda, {\it {Brane embeddings in
  AdS$_4\times\mathbb{CP}^3$}},  {\em Int. J. Mod. Phys.} {\bf A26} (2011)
  2377, [\href{http://arxiv.org/abs/0909.3061}{{\tt arXiv:0909.3061}}].

\bibitem{BabelonBernardTalon03}
O.~Babelon, D.~Bernard, and M.~Talon, {\em {Introduction to classical
  integrable systems}}.
\newblock Cambridge Monographs on Mathematical Physics. Cambridge University
  Press, 2003.

\bibitem{FujitaLiRyuTakayanagi09}
M.~Fujita, W.~Li, S.~Ryu, and T.~Takayanagi, {\it {Fractional quantum Hall
  effect via holography: Chern-Simons, edge states, and hierarchy}},  {\em
  JHEP} {\bf \textbf{06}} (2009) 066,
  [\href{http://arxiv.org/abs/0901.0924}{{\tt arXiv:0901.0924}}].

\bibitem{Wang20a}
Y.~Wang, {\it {Taming defects in $ \mathcal{N} $ = 4 super-Yang-Mills}},  {\em
  JHEP} {\bf \textbf{08}} (2020) 021,
  [\href{http://arxiv.org/abs/2003.11016}{{\tt arXiv:2003.11016}}].

\bibitem{KomatsuWang20}
S.~Komatsu and Y.~Wang, {\it {Non-perturbative defect one-point functions in
  planar $\mathcal{N}=4$ super-Yang-Mills}},  {\em Nucl. Phys.} {\bf
  \textbf{B958}} (2020) 115120, [\href{http://arxiv.org/abs/2004.09514}{{\tt
  arXiv:2004.09514}}].

\bibitem{BissiKristjansenYoungZoubos11}
A.~Bissi, C.~Kristjansen, D.~Young, and K.~Zoubos, {\it {Holographic
  three-point functions of giant gravitons}},  {\em JHEP} {\bf \textbf{06}}
  (2011) 085, [\href{http://arxiv.org/abs/1103.4079}{{\tt arXiv:1103.4079}}].

\bibitem{JiangKomatsuVescovi19a}
Y.~Jiang, S.~Komatsu, and E.~Vescovi, {\it {Structure constants in $
  \mathcal{N} $ = 4 SYM at finite coupling as worldsheet g-function}},  {\em
  JHEP} {\bf 07} (2020) 037, [\href{http://arxiv.org/abs/1906.07733}{{\tt
  arXiv:1906.07733}}].

\bibitem{JiangKomatsuVescovi19b}
Y.~Jiang, S.~Komatsu, and E.~Vescovi, {\it {Exact three-point functions of
  determinant operators in planar $N=4$ supersymmetric Yang-Mills theory}},
  {\em Phys. Rev. Lett.} {\bf \textbf{123}} (2019) 191601,
  [\href{http://arxiv.org/abs/1907.11242}{{\tt arXiv:1907.11242}}].

\bibitem{HiranoKristjansenYoung12}
S.~Hirano, C.~Kristjansen, and D.~Young, {\it {Giant gravitons on AdS$_4 \times
  \mathbb{CP}^3$ and their holographic three-point functions}},  {\em JHEP}
  {\bf \textbf{07}} (2012) 006, [\href{http://arxiv.org/abs/1205.1959}{{\tt
  arXiv:1205.1959}}].

\bibitem{YangJiangKomatsuWu21a}
P.~Yang, Y.~Jiang, S.~Komatsu, and J.-B. Wu, {\it {Three-point functions in
  ABJM and Bethe Ansatz}},  {\em JHEP} {\bf 01} (2022) 002,
  [\href{http://arxiv.org/abs/2103.15840}{{\tt arXiv:2103.15840}}].

\bibitem{YangJiangKomatsuWu21b}
P.~Yang, Y.~Jiang, S.~Komatsu, and J.-B. Wu, {\it {D-branes and orbit
  average}},  {\em SciPost Phys.} {\bf \textbf{12}} (2022) 055,
  [\href{http://arxiv.org/abs/2103.16580}{{\tt arXiv:2103.16580}}].

\end{thebibliography}\endgroup

\end{document}